\documentclass[reprint,amsmath,amssymb,aip]{revtex4-2}
\usepackage{graphicx}
\usepackage{dcolumn}
\usepackage{bm}
\usepackage{physics}
\usepackage{siunitx}
\usepackage{url}
\AtBeginDocument{\RenewCommandCopy\qty\SI} 

\begin{document}

\title{Simplified Shielded MEG-MRI Multimodal System with Scalar-mode Optically Pumped Magnetometers as MEG Sensors}

\author{Yosuke Ito}
\email{yito@kuee.kyoto-u.ac.jp}
\affiliation{Department of Electrical Engineering, Graduate School of Engineering, Kyoto University, Kyoto-daigaku Katsura, Nishikyo-ku, Kyoto 615-8510, Japan.}

\author{Hiroyuki Ueda}
\affiliation{Department of Electrical Engineering, Graduate School of Engineering, Kyoto University, Kyoto-daigaku Katsura, Nishikyo-ku, Kyoto 615-8510, Japan.}

\author{Takenori Oida}
\affiliation{Central Research Laboratory, Hamamatsu Photonics K.K., Japan.}

\author{Takahiro Moriya}
\affiliation{Central Research Laboratory, Hamamatsu Photonics K.K., Japan.}

\author{Akinori Saito}
\affiliation{Central Research Laboratory, Hamamatsu Photonics K.K., Japan.}

\author{Motohiro Suyama}
\affiliation{Global Strategic Challenge Center, Hamamatsu Photonics K.K., Japan.}

\date{\today}

\begin{abstract}

Magnetoencephalography (MEG) conventionally operates within high-performance magnetic shields due to the extremely weak magnetic field signals from the measured objects and the narrow dynamic range of the magnetic sensors employed for detection. This limitation results in elevated equipment costs and restricted usage. Additionally, the information obtained from MEG is functional images, and to analyze from which part of the brain the signals are coming, it is necessary to capture morphological images separately. When MEG and morphological imaging devices are separate, despite their individual high measurement accuracies, discrepancies in positional information may arise. In response, we have developed a low-field magnetic resonance imaging system that incorporates scalar-mode optically pumped magnetometers with a wide dynamic range and exceptionally high measurement sensitivity as sensors for MEG. Operating at low magnetic fields eliminates the need for superconducting coils in magnetic resonance imaging and the high-performance magnetic shields essential for MEG, promising a substantial cost reduction compared to traditional approaches.
We achieved a noise level of about \SI[per-mode=symbol]{16}{\pico\tesla\per\hertz^{1/2}} with a single channel magnetometer, and reached a noise level of \SI[per-mode=symbol]{367}{\femto\tesla\per\centi\metre\per\hertz^{1/2}} through differential measurements. The system successfully conducted MR imaging on a phantom, demonstrating the potential of MEG and MRI fusion. 
\end{abstract}

\maketitle

\section{Introduction}

Optically pumped magnetometers (OPMs) have garnered attention due to its ability to achieve magnetic-field sensitivity comparable to superconducting quantum interference devices (SQUIDs) without the need for cryogenic substances\cite{AllredLymanKornackRomalis2002, KominisKornackAllredRomalis2003}. In recent years, particularly in the field of biomagnetic measurements such as magnetoencephalography (MEG) and magnetocardiography (MCG), research in this area has flourished, partly due to the global helium shortage\cite{BotoMeyerShahAlemKnappeKrugerFromholdLimGloverMorrisBowtellBarnesBrookes2017, BotoHolmesLeggettRobertsShahMeyerMunozMullingerTierneyBestmannBarnesBowtellBrookes2018, HillBotoReaHolmesLeggettColesPapastavrouEvertonHuntSimsOsborneShahBowtellBrookes2020, KimSavukovNewman2019, SanderJodkoWladzinskaHartwigBruehlMiddelmann2020, AlemHughesBuardCheungMaydewGriesshammerHollowayParkLechugaCoolidgeGerginovQuiggSeamesKronbergTealeKnappe2023}. The OPMs' cryogen-free operation enables closer proximity to the signal sources, offering the added benefit of acquiring more substantial signals when compared to SQUID-based magnetometers.  Boto {\it et al}. have demonstrated success in acquiring magnetic signals caused by voluntary movements within a magnetic shield using the OPMs along with meticulously calculated compensating coils\cite{BotoHolmesLeggettRobertsShahMeyerMunozMullingerTierneyBestmannBarnesBowtellBrookes2018}. Furthermore, Limes and colleagues have successfully measured auditory evoked magnetic fields in a favorable magnetic environment in outdoor suburban settings by differentially measuring the precession frequencies of rubidium spin polarization induced by a powerful pulsed laser\cite{LimesFoleyKornackCaligaMcBrideBraunLeeLuciveroRomalis2020}. This type of OPMs is known as scalar-mode OPMs, and by employing them, highly sensitive magnetic field measurements can be achieved even in unshielded environments\cite{ShengLiDuralRomalis2013, ClancyGerginovAlemBeckerKnappe2021}. Utilizing the commercial scalar-mode OPMs, Jaufenthaler {\it et al}. achieved successful observations of the relaxation of magnetic moments in magnetic nanoparticles, even in an unshielded environments\cite{JaufenthalerKornackLebedevLimesKorberLieblBaumgarten2021}.

MEG and other biomagnetic field measurements provide functional images, and this holds true for the measuremnents with OPMs as well. To perform source localization, it is imperative to co-register it with structural imaging\cite{ZetterIivanainenStenroosParkkonen2018, YangWangLiuWangHanJiaPangXieYuZhangXiangNing2023}. However, imaging modalities like computer tomography (CT) and magnetic resonance imaging (MRI), which yield structural data, cannot be acquired simultaneously with MEG, and in fact, it is not feasible to obtain these data using the same device. CT requires X-ray exposure around the subject, which can introduce radiation risks and limit sensor placement. Furthermore, MRI can disrupt sensor operation due to the presence of strong static magnetic fields, gradient fields, and RF fields. Thus, developing a multimodal system that combines MEG with structural imaging devices has been a complex endeavor.

In recent years, there have been proposals for MRI systems utilizing OPM technology\cite{SavukovKaraulanovWurdenSchultz2013, HilschenzItoNatsukawaOidaYamamotoKobayashi2017, KimSavukov2020, HoriOidaMoriyaSaitoSuyamaKobayashi2022}, but these have not been explored as multimodal systems alongside MEG. An advantage of OPM-MEG is the ability to customize a flexible sensor array to match the size and shape of a subject's head\cite{HillBotoReaHolmesLeggettColesPapastavrouEvertonHuntSimsOsborneShahBowtellBrookes2020}. However, this advantage makes the process of overlaying functional and anatomical images, typically done more easily in conventional SQUID-MEG, somewhat challenging. Thus, there is a growing interest in achieving a multimodal system for OPM-MEG that can also capture anatomical images.

In addressing this issue, we propose a multimodal system for MEG by differential measurements with scalar-mode OPMs and low-field (7-mT) MRI with a non-cryogenic pickup coil and normal conducting coil sets in a simplified magnetic shield. In comparison to using OPM as an MRI detector, pickup coils exhibit superior characteristics in terms of receiving bandwidth and stability. Therefore, in this study, OPM was employed for MEG detection, while the pickup coil was utilized for MRI detection. Frequency range of scalar-mode OPMs is determined by the pump-probe cycle, making it capable of measuring signals up to kHz\cite{LiLiuJinTettehAkitiDaiXuEricTheophilusNwodom2022}. Moreover, the scalar-mode OPMs, characterized by their wide dynamic range, facilitate high-sensitivity measurements even within a simplified magnetic shielding environment.
On the other hand, in the case of MRI, the economical normal conduction coil sets allow for easy switching of the static magnetic fields, ensuring that it does not hinder the operation of the scalar-mode OPMs during MEG measurements. In our developed 7-mT MRI system, Larmor frequency is around 300 kHz, providing sufficient detection sensitivity even with the non-cryogenic pickup coil\cite{SavukovSeltzerRomalis2007}.

This study describes the development of a scalar-mode OPM module enabling differential measurements and its integration into a simplified magnetic shielded low-field MRI system, including experimental assessments of its performance.

\begin{figure*}
 \centering
 \includegraphics[width=.9\linewidth]{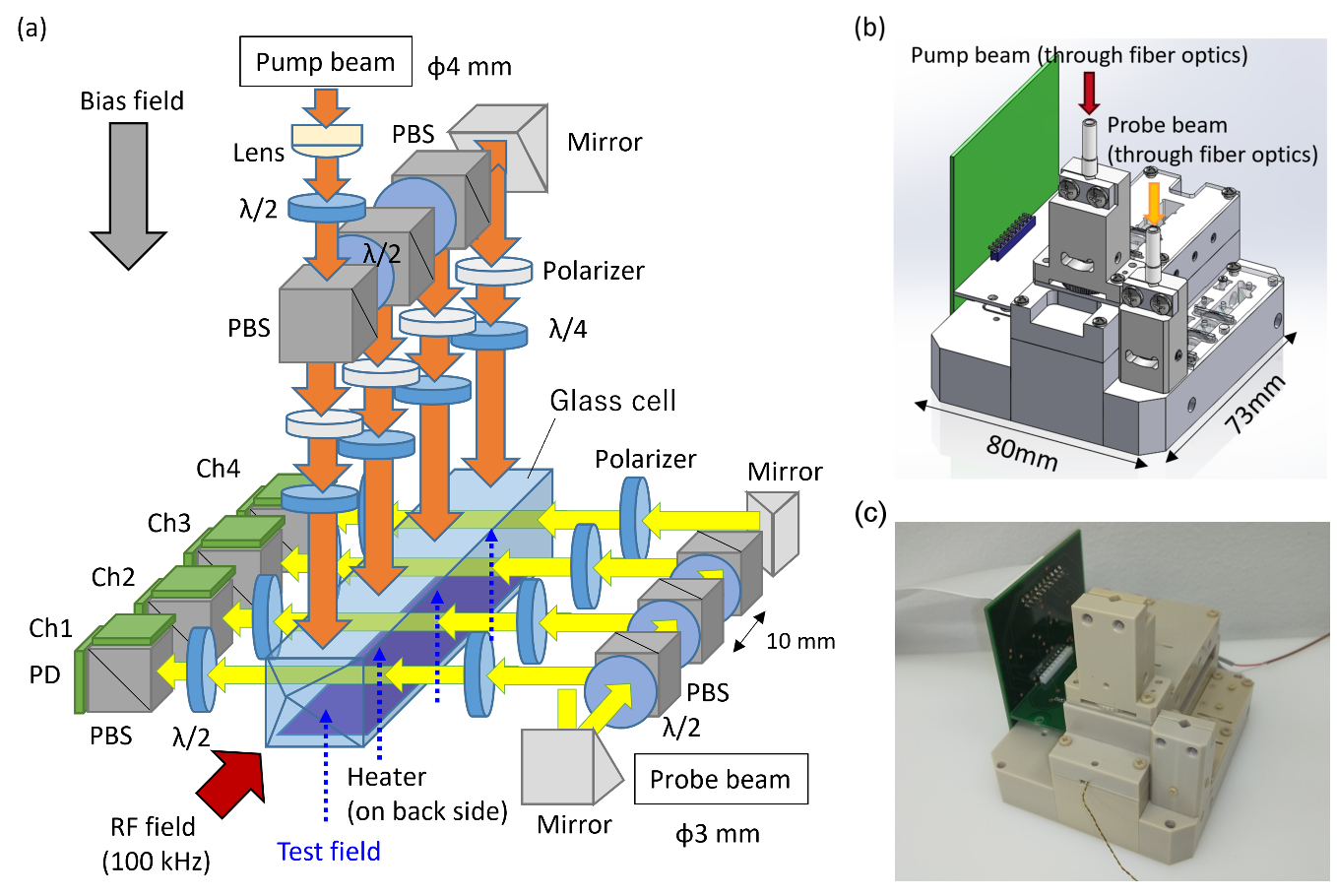}

 \caption{4-ch scalar-mode OPM module. (a) schematic drawing, (b) overview, and (c) photograph of 4-ch scalar-mode OPM module. The pump and probe beams were introduced to the module via fiber optics, undergo polarization adjustment by waveplates, and then were distributed to four points within a glass cell by polarizing beam splitters (PBSs). The beams passing through the glass cell were detected by polarimeters consisting of a PBS and two photodiodes (PDs). The glass cell, which contains potassium, helium, and nitrogen, was heated by an electric heater installed on one side of the glass cell.} 
\label{124533_9Nov23}
\end{figure*}

\begin{figure*}
 \centering
 \includegraphics[width=.9\linewidth]{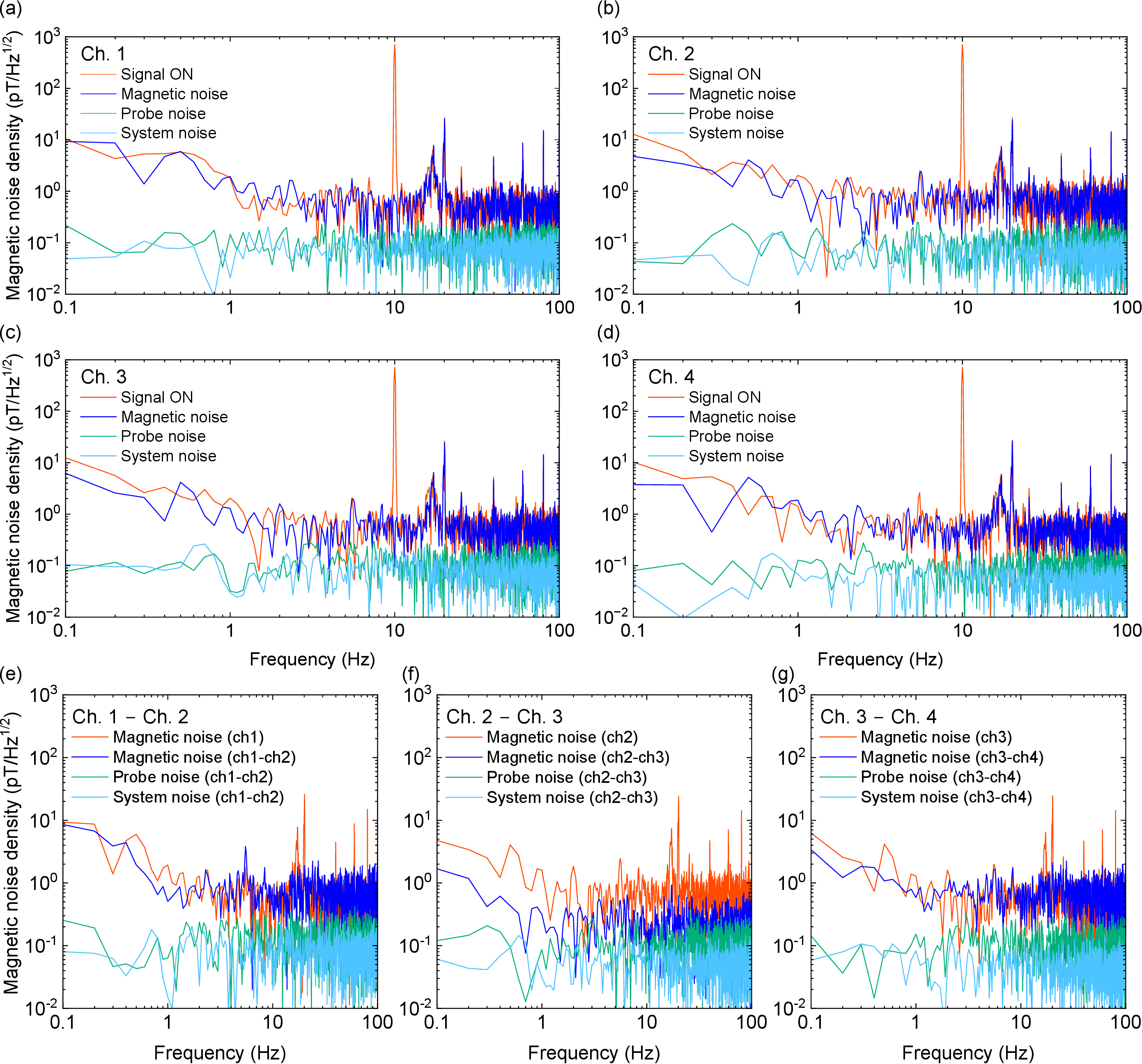}

 \caption{Potential of the 4-ch scalar-mode OPM module. Magnetic noise denisty at (a) Ch. 1, (b) Ch. 2, (c) Ch. 3 and (d) Ch. 4 of 4-ch scalar-mode OPM module in magnetic shield. The reference field was 10-Hz sinusoidal wave with an amplitude of 318 pT. Probe noise and system noise were estimated from noise level of probe beam noise and amplifier noise. Magnetic noise denisty based on differential signals of each pair of two channels (e) Ch. 1 $-$ Ch. 2, (f) Ch. 2 $-$ Ch. 3 and (g) Ch. 3 $-$ Ch. 4 of 4-ch scalar-mode OPM module in magnetic shield.} 
\label{132720_28Nov23}
\end{figure*}

\begin{figure*}
 \centering
 \includegraphics[width=.9\linewidth]{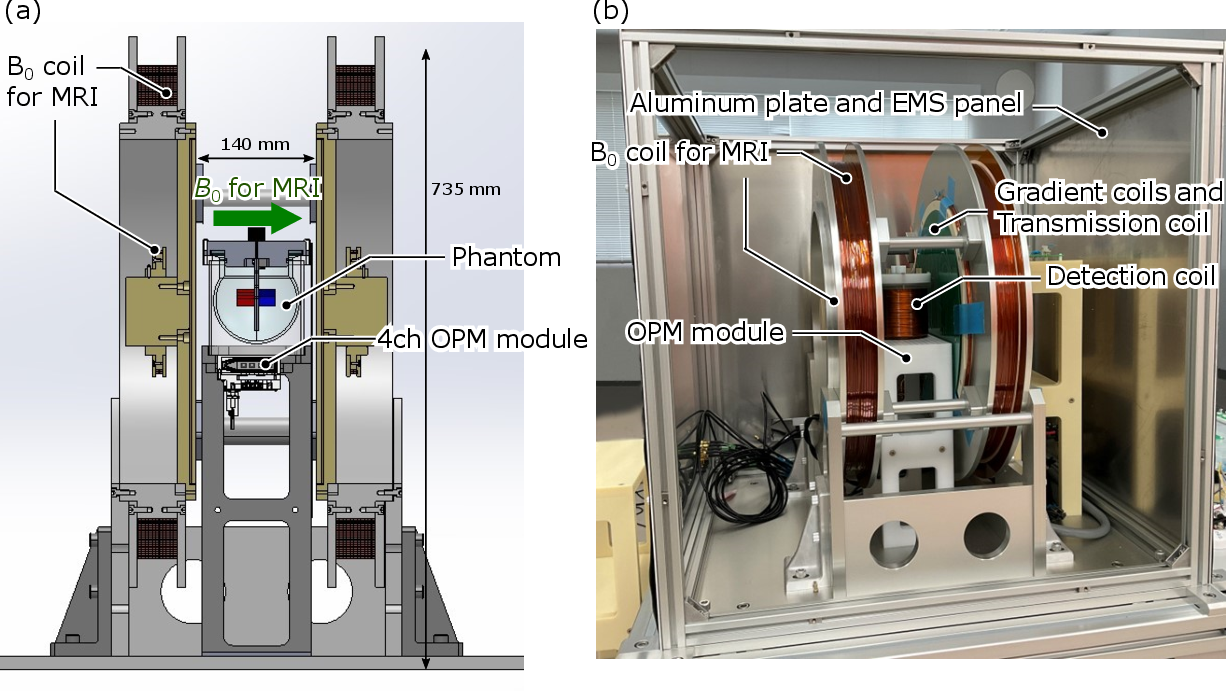}

 \caption{Developed 7-mT MRI system with the 4-ch scalar-mode OPM module. (a) schematic drawing and (b) photograph of MRI system. As illustrated in (b), The $B_0$ coil for MRI consists of four coils to ensure the uniformity of the static magnetic field. The 4-ch scalar-mode OPM module was placed under the phantom, which was surrounded by a detection coil, in the coil set consisting of a transmission coil and gradient coils. The coil set was also present on the opposing surfaces, with the phantom sandwiched in between. The MRI system was surrounded by aluminum plates and EMS panels.} 
\label{023824_28Nov23}
\end{figure*}

\begin{figure*}
 \centering
 \includegraphics[width=.9\linewidth]{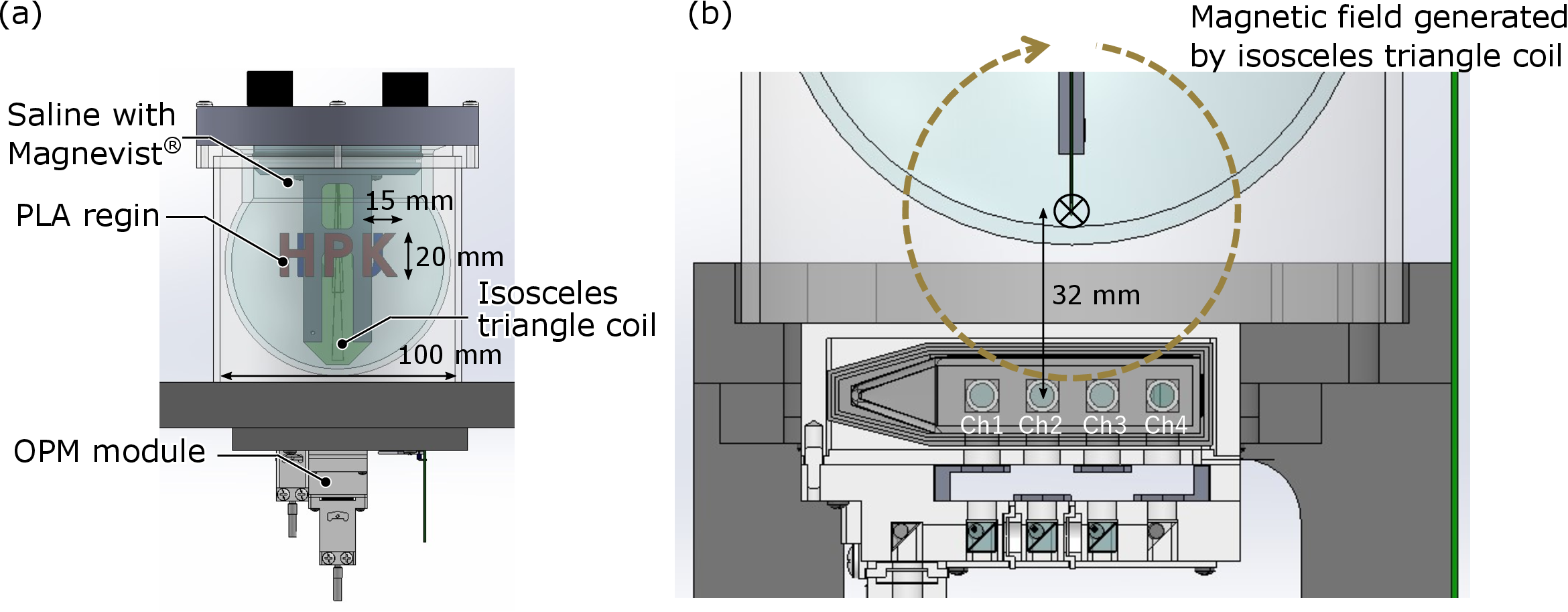}

 \caption{Phantom with objects of 'HPK' and 'KU', and an isosceles triangle coil. (a) schematic drawing of phantom utilized in the measurements and (b) relationship in position between magnetic field generated by the isosceles triangle coil and the OPM module.} 
\label{033808_28Nov23}
\end{figure*}

\section{Methods}

\subsection{Scalar-mode optically pumped magnetometer module}

Figure \ref{124533_9Nov23}(a) illustrates a schematic diagram of the 4-ch scalar-mode OPM module that we have developed. The pump and probe beams were introduced to the module via fiber optics, undergo polarization adjustment by waveplates, and then were distributed to four points within a glass cell by polarizing beam splitters (PBSs). The diameter of the pump beam was 4 mm, and that of the probe beam was 3 mm. The beams passing through the glass cell were detected by polarimeters consisting of a PBS and two photodiodes (PDs). The glass cell, which contains potassium, helium, and nitrogen, was heated by an electric heater installed on one side of the glass cell. The overview of the module is shown in Fig.~\ref{124533_9Nov23}(b). The footprint of the module was $80$ mm $\times 73$ mm, within which four channels were arranged linearly. Each channel was positioned at 10 mm apart, allowing for the calculation of tangential differential outputs by subtracting their respective signals. Figure \ref{124533_9Nov23}(c) displays a photograph of the module. The module is fabricated from polyetheretherketone (PEEK). The wire for the heater was introduced from the opposite side of the fiber optics ports.

The magnetic noise density at each channel in a magnetic shield is shown in Fig.~\ref{132720_28Nov23}(a), (b), (c) and (d). The shielding factor of the magnetic shield was $>10^4$ at \SI{10}{\hertz}. We applied spatially uniform 10-Hz sinusoidal wave as a reference field. The measurement time was 10 s and the sampling rate was \SI{200}{\hertz}. The magnetic noise density was 560, 669, 580, and \SI[per-mode=symbol]{562}{\femto\tesla\per\hertz^{1/2}} at Ch. 1, 2, 3, and 4, respectively. The estimated probe beam noise was 69, 55, 143, and \SI[per-mode=symbol]{102}{\femto\tesla\per\hertz^{1/2}} at Ch. 1, 2, 3, and 4, respectively. The probe beam noise was less than the magnetic field noise, therefore the magnetic noise still remained.
Figure \ref{132720_28Nov23}(e), (f) and (g) shows magnetic noise density based on differential signals of each pair of two channels. The magnetic noise density was \SI[per-mode=symbol]{560}{\femto\tesla\per\centi\metre\per\hertz^{1/2}} at Ch.1 $-$ Ch.2, \SI[per-mode=symbol]{154}{\femto\tesla\per\centi\metre\per\hertz^{1/2}} at Ch.2 $-$ Ch.3, and \SI[per-mode=symbol]{564}{\femto\tesla\per\centi\metre\per\hertz^{1/2}} at Ch.3 $-$ Ch.4. 
For Ch. 2 $-$ Ch. 3, there was a substantial enhancement in measurement sensitivity, whereas there was less apparent improvement in other cases. This could be attributed to the suppression of the impact of non-uniform magnetic fields in the central region of the OPM module. Conversely, as one deviated from the central region, the impact of non-uniform environmental magnetic fields became more prominent.

\subsection{MRI system}

Figure \ref{023824_28Nov23}(a) depicts a schematic diagram of an MRI apparatus with the 4-ch scalar-mode OPM module developed by our group. The OPM module was positioned beneath the phantom, which was surrounded by a detection coil, enclosed between the coils for the static magnetic field and the coil set for the transmission and gradient magnetic fields. The strength of the static field for MRI was 7 mT.
The photograph of this system is presented in Fig.~\ref{023824_28Nov23}(b). The entire system was enclosed by aluminum plates and 5-layer EMS panels (Medical-aid Co., Ltd.), and the effect of the EMS panels reduces magnetic noise in the frequency range below 100 Hz to approximately 1/5. While the shielding performance is about 1/2000 compared to shields commonly used for MEG, allows for cost-effective production.

For MRI imaging, a spin-echo sequence was employed, and the imaging parameters are detailed in Table \ref{183338_28Nov23}.
The scan time at NEX = 1 was approximately 5 minutes, while at NEX = 8, it was about 41 minutes, and at NEX = 16, it was approximately 82 minutes.

\begin{table}
\centering
\caption{MR imaging parameters employed in experiments.}
\label{183338_28Nov23}
 \begin{tabular}{cc}
\hline
  Prameter& Value\\
\hline
 FoV& $96\times 96\times 96$ \\
  Matrix& $32\times 32\times 32$\\
  TR& 300 ms \\
  TE& 25 ms \\
  FA& \ang{90} \\
  BW& 100 Hz/pixel\\
  NEX& 1, 8, 16\\
\hline
\end{tabular}
\label{041359_28Nov23}
\end{table}

\subsection{Phantom}

Figure \ref{033808_28Nov23}(a) illustrates an illustration of the phantom used in the measurements for the OPM performance. The phantom was filled with saline solution supplemented with magnevist$^\text{\textregistered}$ with 1 mM, and contained objects with adhered labels 'HPK' and 'KU' made of PLA resin. The size of each character was 20 mm in height and 15 mm in width. The T1 and T2 relaxation time were 94 ms and 191 ms, respectively. Additionally, an isosceles triangle coil was positioned beneath the phantom to align with its base. As depicted in Fig.~\ref{033808_28Nov23}(b), the distance between the base of the coil and the measurement region of the OPM module was 32 mm, ensuring that the coil was placed directly above channels 2 and 3 of the OPM module.

\section{Results}

\begin{figure*}
 \centering
 \includegraphics[width=.9\linewidth]{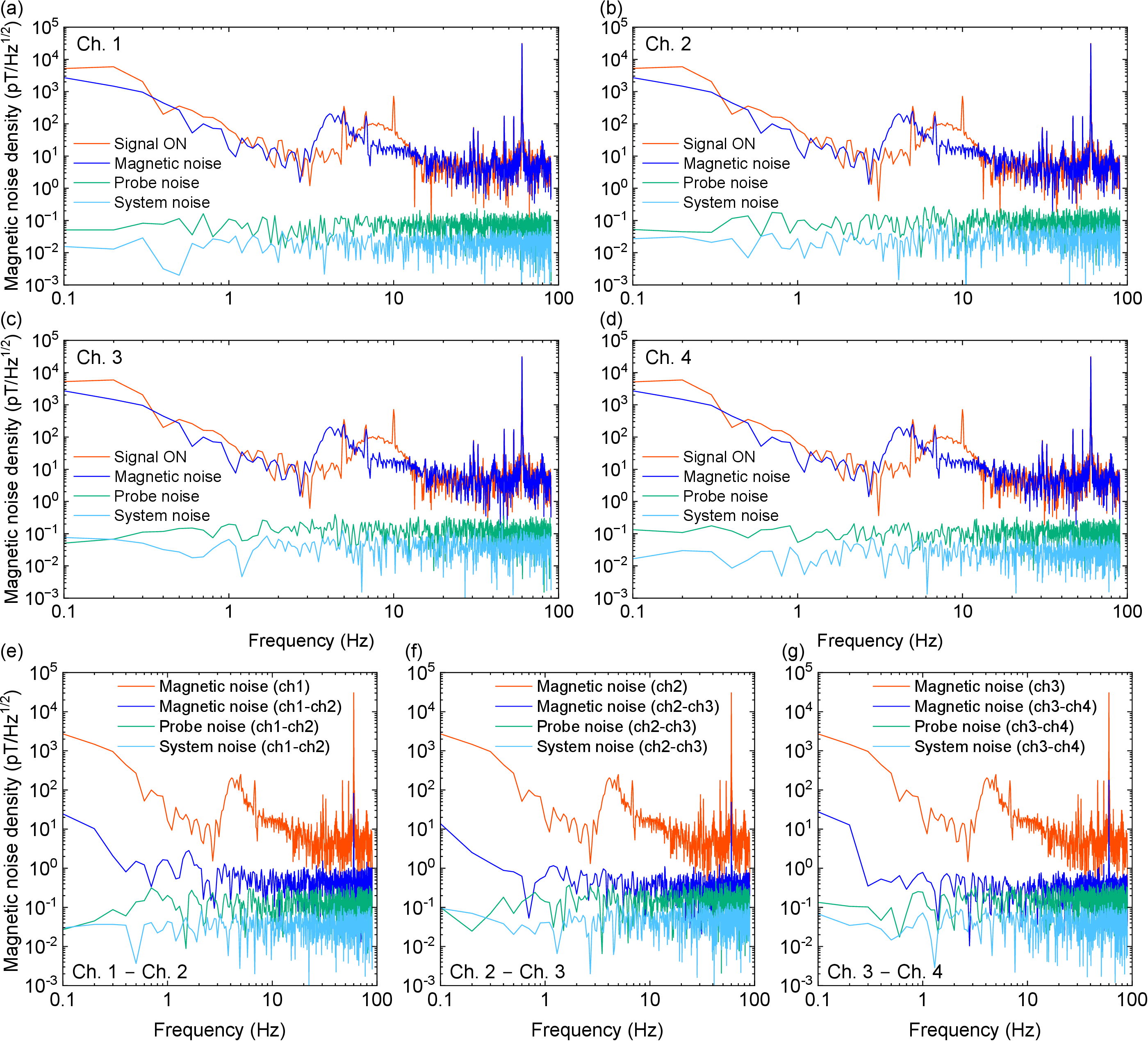}

 \caption{Magnetic noise measurements with the scalar-mode OPM installed in the MRI system. Magneitc noise density at (a) Ch. 1, (b) Ch. 2, (c) Ch. 3 and (d) Ch. 4 of 4-ch scalar-mode OPM module. The reference field was 10-Hz sinusoidal wave with an amplitude of 318 pT. Probe noise and system noise were estimated from noise level of probe beam noise and amplifier noise. Magneitc noise density based on the differential signals of each pair of two channels (e) Ch. 1 $-$ Ch. 2, (f) Ch. 2 $-$ Ch. 3 and (g) Ch. 3 $-$ Ch. 4 of 4-ch scalar-mode OPM module. Since the sensitivity direction of the OPM module was perpendicular to the arrangement of channels, differentials along the tangential direction of the phantom can be obtained.} 
\label{042816_28Nov23}
\end{figure*}

\begin{figure*}
 \centering
 \includegraphics[width=.9\linewidth]{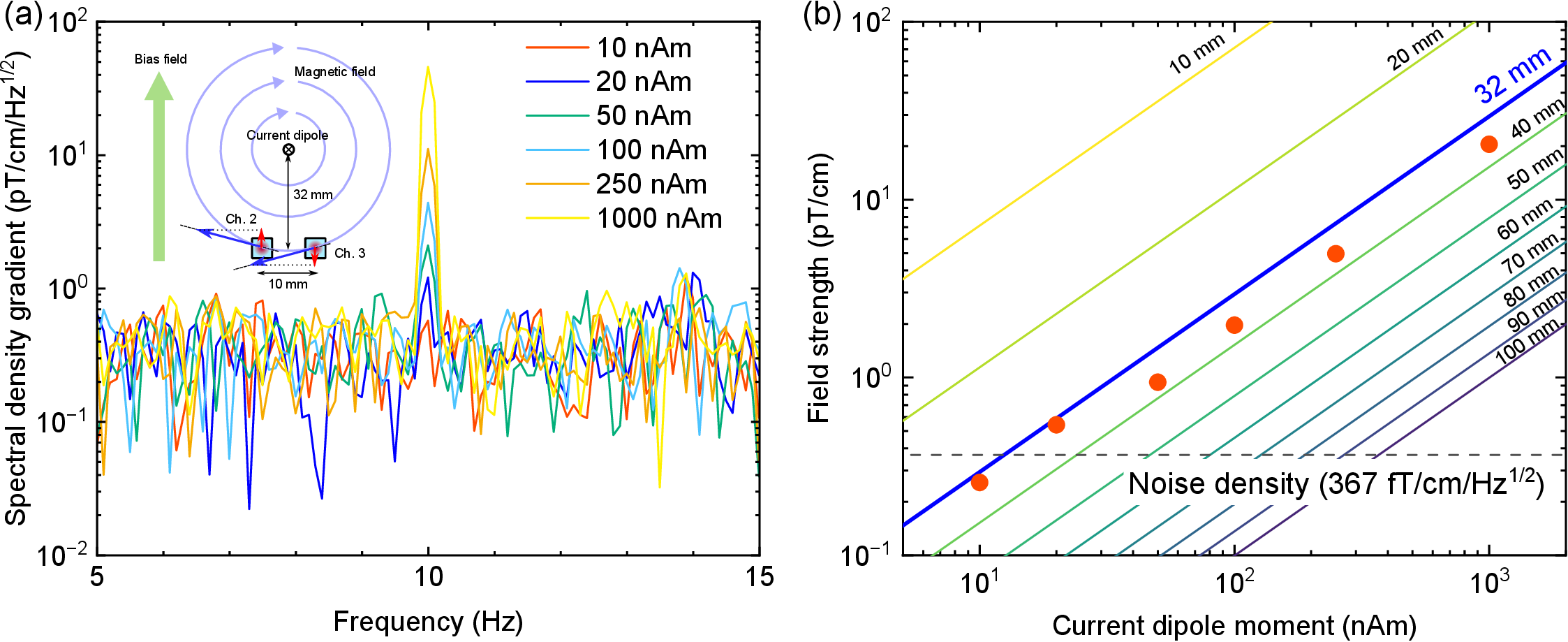}

 \caption{Current dipole moment measurements with the scalar-mode OPM. (a) FFT spectra of differential measurements at Ch.2 $-$ Ch.3 for the magnetic field generated by the isosceles triangle coil. A sinusoidal current of 10 Hz was applied to the isosceles triangle coil, demonstrating measurements for current dipole magnitudes of 1000 nAm, 250 nAm, 100 nAm, 50 nAm, 20 nAm, and 10 nAm formed by its base. (b) Field strength as a function of current dipole moment. The background solid lines incdicate field strength depending on distance between the current dipole and sensing region calculated by Biot-Savart law.} 
\label{185330_28Nov23}
\end{figure*}

\begin{figure}
 \centering
 \includegraphics[width=.9\linewidth]{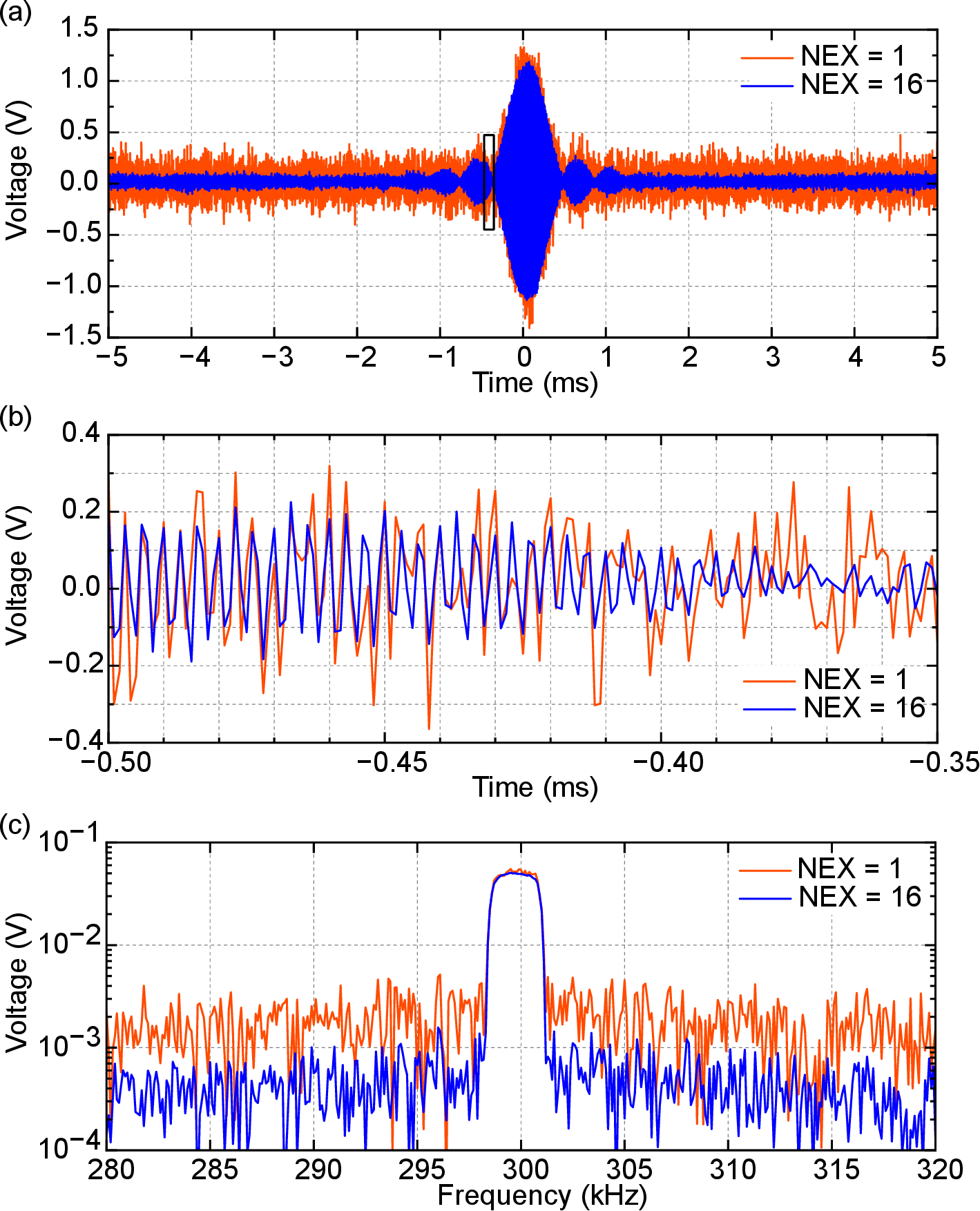}

 \caption{Spin echo signals of the MRI system. (a) Time evolution of spin echo signals from a phantom with an absence of internal objects in the absence of phase encoding and slice encoding with NEX = 1 and NEX = 16. (b) An enlarged view of the black frame in (a), where amplitude-modulated signals at approximately 300 kHz were observed. (c) FFT spectra of the spin echo signals.} 
\label{033329_29Nov23}
\end{figure}

\begin{figure}
 \centering
 \includegraphics[width=.9\linewidth]{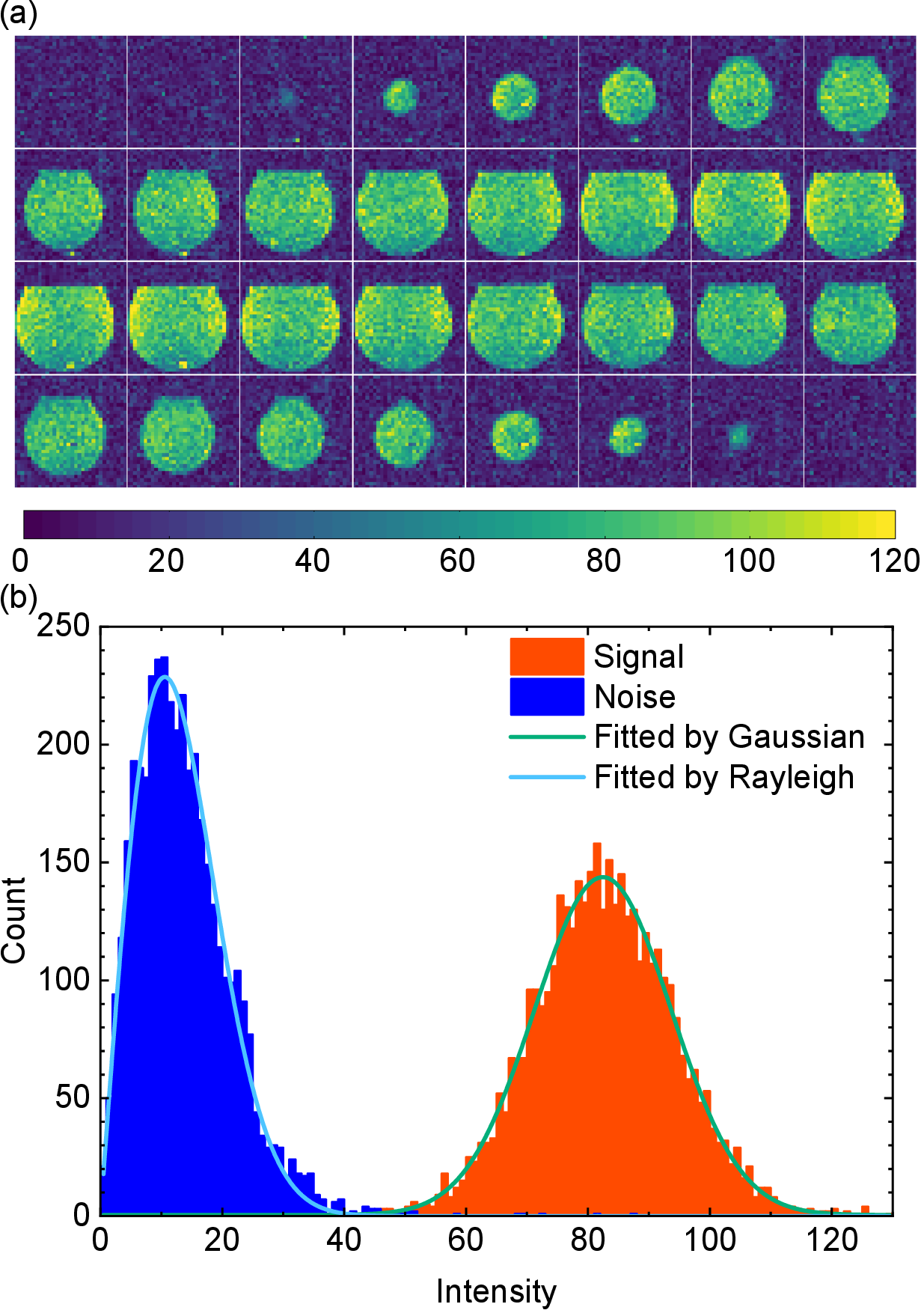}

 \caption{Evaluation of images with the MRI system. (a) Reconstruction images of a phantom with an absence of internal objects. (b) Histograms of signal and noise domains. Signal and noise domains were defined as the central region of the phantom ($16\times 16\times 16$ voxels) and  $8\times 8\times 8$ voxels at each of the 8 corners in the imaging region. The solid lines represent distribution curves of histrograms of signal and noise domains fitted by Gaussian and Rayleigh distribution functions, respectively.} 
\label{050845_29Nov23}
\end{figure}

\begin{figure*}
 \centering
 \includegraphics[width=.9\linewidth]{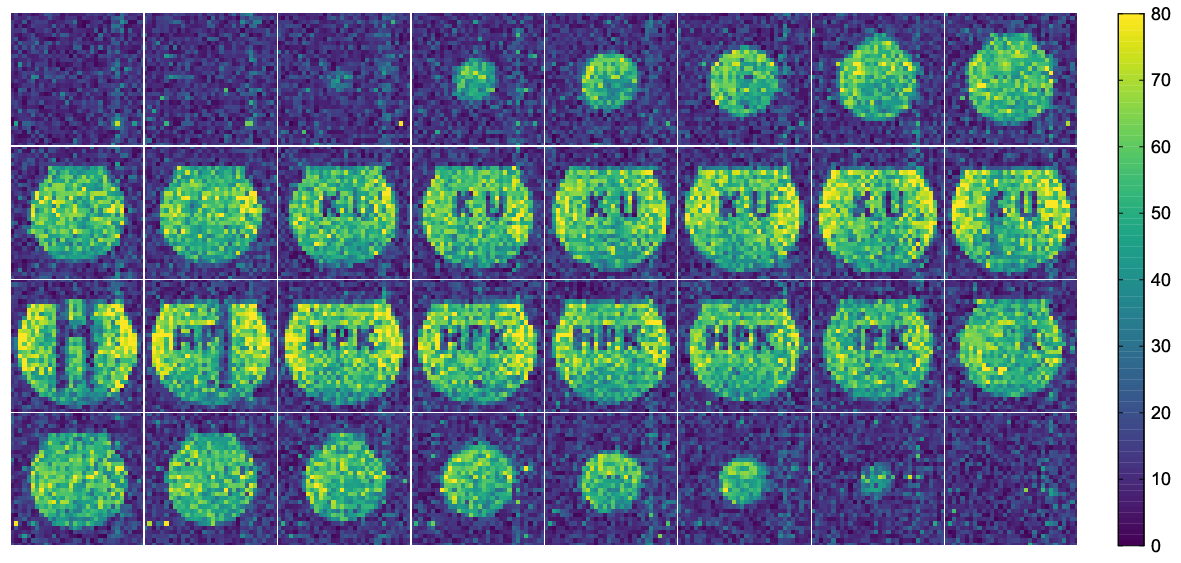}

 \caption{Reconstruction images of 32 different slices. The objects 'HPK' and 'KU' were observed on different slices.} 
\label{040919_29Nov23}
\end{figure*}

\subsection{Sensitivity of 4-ch scalar-mode OPM module}

Figure \ref{042816_28Nov23}(a), (b), (c) and (d) shows magnetic noise density of each channel of the 4-ch scalar-mode OPM module. We applied spatially uniform 10-Hz sinusoidal wave with an amplitude of 318 pT as a reference signal. The measurement time was 10 s and the sampling rate was 180 Hz. The magnetic noise levels at 10 Hz were consistently about 16.7 pT/Hz$^{\mathrm{1/2}}$ across all channels. In terms of probe noise derived from probe beam, it varied significantly across channels, measuring 69, 55, 143, and 102 fT/Hz$^{\mathrm{1/2}}$ for channels 1, 2, 3, and 4, respectively. This variation was attributed to the uneven intensity of probe beam in each channel and the distinct paths taken by each probe beam. Furthermore, the system noise, calculated from amplifier noise, measured 24, 22, 54, and 34 fT/Hz$^{\mathrm{1/2}}$ for channels 1, 2, 3, and 4, respectively, with channel 3 displaying slightly elevated values, but overall comparable magnitudes.

Figure \ref{042816_28Nov23}(e), (f) and (g) presents magnetic noise density of the outputs obtained by differencing each pair of two channels. Since the sensitivity direction of the OPM module aligns with the vertical direction of the channel arrangement, taking the difference of each channel results in measuring the tangential difference as the vertical component along the channel array.
Magnetic noise was effectively subtracted by the differential measurements. The magnetic noise was \SI[per-mode=symbol]{538}{\femto\tesla\per\centi\metre\per\hertz^{1/2}} at Ch.1 $-$ Ch.2, \SI[per-mode=symbol]{367}{\femto\tesla\per\centi\metre\per\hertz^{1/2}} at Ch.2 $-$ Ch.3, and \SI[per-mode=symbol]{261}{\femto\tesla\per\centi\metre\per\hertz^{1/2}} at Ch.3 $-$ Ch.4.
The noise level at Ch.1 $-$ Ch.2 closely resembled the values measured within the magnetic shield. This suggests that even with a simple shield, the impact of spatially uniform environmental magnetic fields can be adequately mitigated through differential measurements.

Figure \ref{185330_28Nov23}(a) shows FFT spectra of differential measurements at Ch.2 $-$ Ch.3 for the magnetic field generated by the isosceles triangle coil. A sinusoidal current of 10 Hz was applied to the isosceles triangle coil. Regarding its base as the current dipole, the magnitude was varied from 1000 to 10 nAm. The peak at 10 Hz can be observed from 1000 to 20 nAm, but not observed with 10 nAm. In this region, the magnitude of the lateral magnetic field was significantly smaller compared to the bias field, and it also deviated considerably from the resonance frequency of the OPM, thus requiring consideration only of the influence of the vertical magnetic field. Therefore, the measurement results were presumed to represent the difference in the vertical magnetic field for each channel.

Figure \ref{185330_28Nov23}(b) shows field strength as a function of a current dipole moment. In our system, noise density was estimated as \SI[per-mode=symbol]{367}{\femto\tesla\per\centi\metre\per\Hz^{1/2}}, therefore the detectable current dipole moment was about 15 nAm. The backgound solid lines indicated field strength of the verical magnetic fields at the OPM module depending on distance between the current dipole and sensing region calculated by Biot-Savart law. The experimental data closely matched the calculated results for a scenario where the current dipole and sensing area were separated by 32 mm; however, the magnetic field intensity was consistently lower than expected along the 32-mm line. This discrepancy was likely due to the actual distance between the current dipole and the sensing area being a few millimeters greater than 32 mm. Additionally, when measuring current dipole moment of 100 nAm with this OPM, it is anticipated that measurements can be taken up to approximately 60 mm away from the current dipole. Given that the distance from the sensing area to the sensor module surface is about 12 millimeters, signal acquisition is estimated to be feasible up to approximately 48 millimeters from the scalp. This result indicated that this OPM module possessed performance capabilities withstanding magnetoencephalography, because the strength of equivalent current dipoles caused by human brain activities is estimated to be 10 -- 100 nAm\cite{HaemaelaeinenHariIlmoniemiKnuutilaLounasmaa1993}.

\subsection{MR imaging}

Figure \ref{033329_29Nov23}(a) shows spin echo signals from a phantom with an absence of internal objects and in the absence of phase and slice encoding with NEX = 1 and NEX = 16. Figure \ref{033329_29Nov23}(b) is an enlarged view of Fig.~\ref{033329_29Nov23}(a) and FFT spectra of spin echo signals are shown in Fig.~\ref{033329_29Nov23}(c). Amplitude-modulated signals at approximately 300 kHz, which is Larmor frequency of the static field of 7 mT, was observed in Fig.~\ref{033329_29Nov23}(b) and (c). The noise level was roughly estimated to be about one-fourth lower in NEX = 16 compared to NEX = 1, indicating the presence of additive effects.

Figure \ref{050845_29Nov23}(a) shows reconstruction images of the phantom with an absence of internal objects acquired by applying phase encoding and slice encoding and imaging parameters in Table \ref{183338_28Nov23}. In this case, we set the NEX to 8.
Figure \ref{050845_29Nov23}(b) shows histograms of signal and noise domains of the reconstructed images at NEX = 8. The signal and noise domains were defined as the central region of the phantom ($16\times 16\times 16$ voxels) and  $8\times 8\times 8$ voxels at each of the 8 corners in the imaging region. 
The histrograms of signal and noise domains were fitted by Gaussian and Rayleigh distribution because it is well known that the image intensity of MR images follows Rician distribution\cite{HaKonSamuel1995}: it can be approximated by Rayleigh distribution in low intensity regions and by Gaussian distribution in high intensity regions. Therefore, the histrograms of signal and noise domains were fitted by Gaussian and Rayleigh distribution.

The mode and the standard deviation (SD) of the intensity at the signal domain were $82.44 \pm 0.12$ and $11.23 \pm 0.14$, and those of the noise domain were $13.23 \pm 0.08$ and $4.53 \pm 0.03$, respectively. 
The values following the $\pm$ symbol represent the standard errors of the fitting.
The signal-to-noise (SNR) calculated as the ratio of the mode of the intensity at the signal domain and the noise domain was $6.23 \pm 0.05$, and that calculated as the average intensity divided by the SD of the signal domain was $7.34 \pm 0.10$. According to Rose model, SNR per pixel should be larger than 5 to discriminate signal from noise\cite{Rose1973}. Considering 99\% confidence interval of the signal domain, the SNR per pixel of the signal domain was larger than 7.8. It was found that artifacts and noise at specific frequencies were quite minimal.

Using the phantom illustrated in Fig.~\ref{033808_28Nov23} as the target object, MR signals were acquired by applying phase encoding and slice encoding and imaging parameters in Table \ref{183338_28Nov23} with NEX = 8. 
The reconstructed images are presented in Fig.~\ref{040919_29Nov23}. The images were obtained in 32 different slices. In Fig.~\ref{033808_28Nov23}, the characters 'KU' were located on the reverse side of the characters 'HPK'. As the slice transitions, a noticeable interchange of these characters appeared.

\section{Discussion}

\begin{figure*}
 \centering
 \includegraphics[width=.9\linewidth]{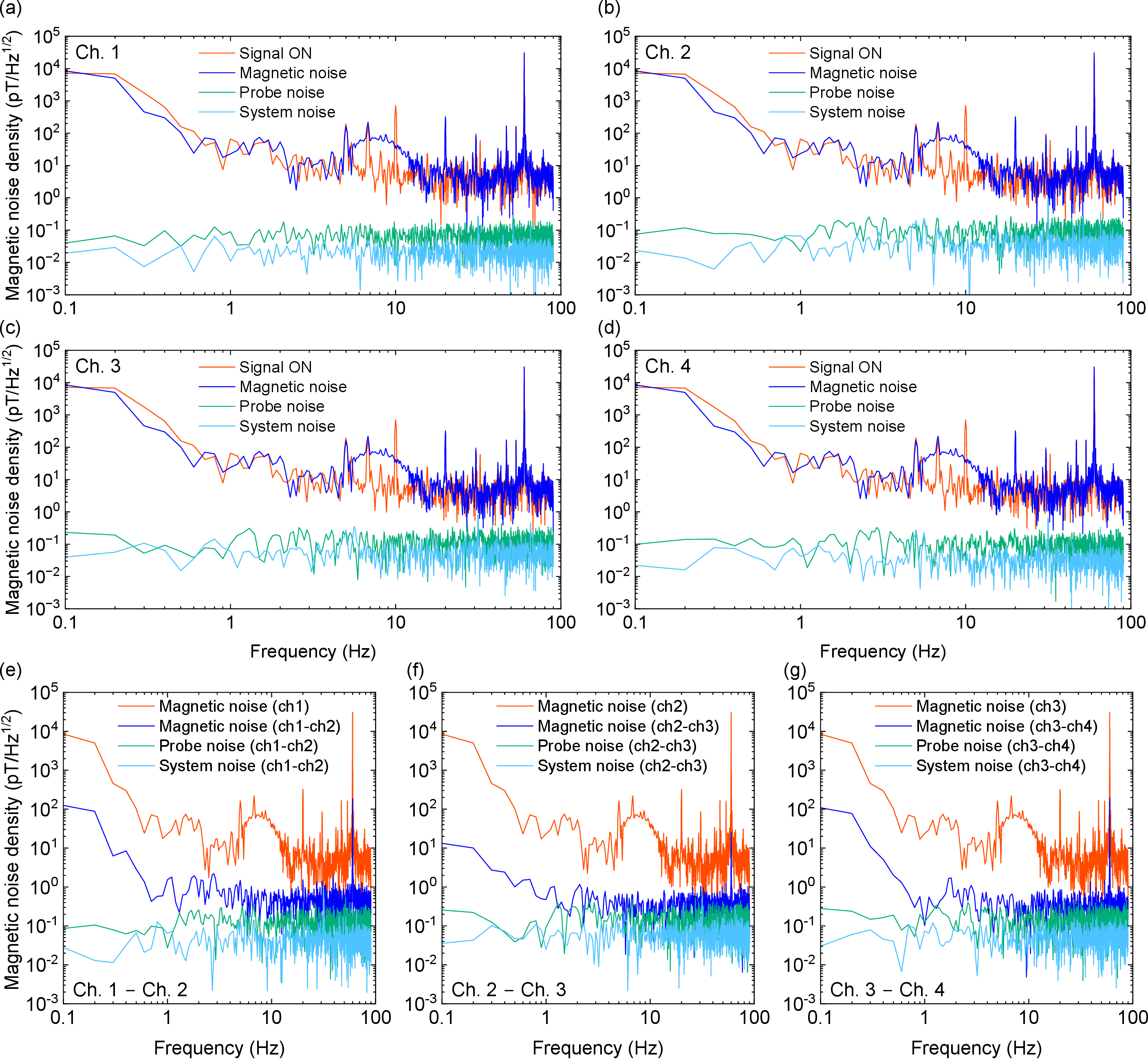}

 \caption{Assessment of impact of the MRI operation on the performance of the scalar-mode OPM. Magnetic noise density at (a) Ch. 1, (b) Ch. 2, (c) Ch. 3 and (d) Ch. 4, and differential measurements of each pair of two channels (e) Ch. 1 $-$ Ch. 2, (f) Ch. 2 $-$ Ch. 3 and (g) Ch. 3 $-$ Ch. 4 of the OPM module after MRI operation.} 
\label{100314_29Nov23}
\end{figure*}

\begin{figure*}
 \centering
 \includegraphics[width=.9\linewidth]{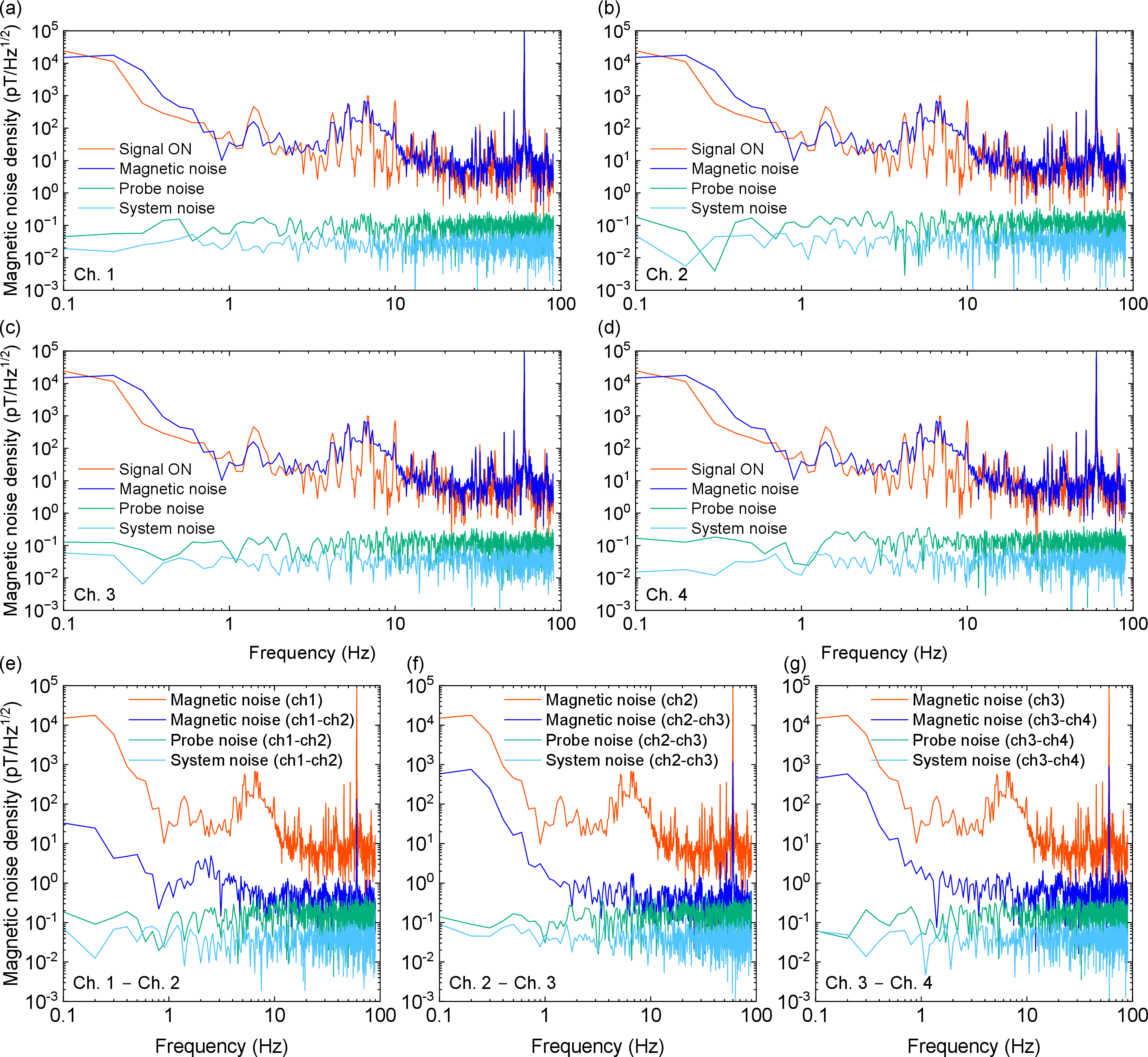}

 \caption{Magnetic Shieldless examination of the scalr-mode OPM. Magnetic noise density at (a) Ch. 1, (b) Ch. 2, (c) Ch. 3 and (d) Ch. 4, and differential measurements of each pair of two channels (e) Ch. 1 $-$ Ch. 2, (f) Ch. 2 $-$ Ch. 3 and (g) Ch. 3 $-$ Ch. 4 of the OPM module without magnetic shield.} 
\label{112714_29Nov23}
\end{figure*}

With our system, we were able to present the potential for MEG measurements and validate MR imaging. In this multimodal system, where MEG and MRI are intended to be sequentially measured, it is crucial to investigate whether the application of a relatively large magnetic field during MRI impacts MEG measurements. Hence, we conducted a measurement of the magnetic field sensitivity of the OPM module after MRI operation, and the results are illustrated in Fig.~\ref{100314_29Nov23}.
The magnetic noise density of each channel after MRI operation was about \SI[per-mode=symbol]{42}{\pico\tesla\per\hertz^{1/2}}, and MRI operation had resulted in an approximately 2.5-fold increase in noise across all channels. However, in the case of differential measurements, there was a consistent improvement in magnetic noise density for all measurements. This suggests that the significant current flow in the shield wall during MRI operation altered the magnetic field environment within the shield. The consistency of probe noise further supports this observation.
Despite fluctuations in the magnetic field environment, the capability of maintaining robust sensitivity through differential measurements positions our newly developed system as a promising MEG and MRI fusion system.

Finally, we examined the operation without any magnetic shields. Figure \ref{112714_29Nov23} illustrates the magnetic noise density when the EMS panels used as an magnetic shield was removed. The magnetic noise density for each channel was approximately \SI[per-mode=symbol]{60}{\pico\tesla\per\hertz^{1/2}}. However, through differential measurements, the  magnetic noise density could be reduced to about \SI[per-mode=symbol]{370}{\femto\tesla\per\centi\metre\per\hertz^{1/2}} for conditions other than Ch. 3 $-$ Ch. 4, demonstrating that operation without the magnetic shield is feasible. Nevertheless, in the frequency range below 5 Hz, the noise difference was not discernible, and under Ch. 3 $-$ Ch. 4 condition, the magnetic noise density was more than double compared to the shielded case. Therefore, the advanced measures, such as higher-order differential measurements are likely required for measurements without magnetic shields.

\section{Conclusion}

We designed and assessed a prototype combining MEG utilizing a scalar-mode OPM module and MRI with non-cryogenic pickup coils, exploring its capabilities. With a simplified magnetic shield, we achieved a noise level of about \SI[per-mode=symbol]{16}{\pico\tesla\per\hertz^{1/2}} with a single channel magnetometer, and reached a noise level of \SI[per-mode=symbol]{367}{\femto\tesla\per\centi\metre\per\hertz^{1/2}} through differential measurements. The system successfully conducted MR imaging on a phantom, demonstrating the potential of MEG and MRI fusion. Furthermore, we demonstrated the sustained noise levels in differential measurements both pre and post MRI operation, showcasing the viability of our system as a fusion device. Future challenges encompass advancing sensor performance through enhancements in lasers and optical systems, along with accelerating imaging via sequence improvements in MRI. Our prospective aim involves utilizing this system for actual MEG measurements, progressing towards the realization of MEG and MRI fusion.

\begin{acknowledgments}
 This work was supported by a Grant-in-Aid for Research (21H03807) from the Ministry of Education, Culture, Sports, Science, and Technology (MEXT), Japan.
\end{acknowledgments}

\end{document}